\documentclass[prd,onecolumn,nofootinbib,showpacs,showkeys]{revtex4}
\usepackage{bm,amsmath,amssymb}

\begin{document}

\title{Thermal fluctuations in Einstein-Cartan-Sciama-Kibble-Dirac bouncing cosmology}
\author{Nikodem J. Pop{\l}awski}
\affiliation{Department of Physics, Indiana University, Swain Hall West, 727 East Third Street, Bloomington, Indiana 47405, USA}
\email{nipoplaw@indiana.edu}
\date{\today}

\begin{abstract}
We study cosmological perturbations arising from thermal fluctuations in the big-bounce cosmology in the Einstein-Cartan-Sciama-Kibble theory of gravity.
We show that such perturbations cannot have a scale-invariant spectrum if fermionic matter minimally coupled to the torsion tensor is macroscopically averaged as a spin fluid, but have a scale-invariant spectrum if the Dirac form of the spin tensor of the fermionic matter is used.
\end{abstract}

\pacs{04.50.Kd, 98.80.Cq}
\keywords{torsion, spinors, minimal coupling, gravitational repulsion, big bounce, nonsingular universe.}

\maketitle

The Einstein-Cartan-Sciama-Kibble (ECSK) theory of gravity naturally extends general relativity by removing its constraint of a symmetric affine connection \cite{KS}.
The antisymmetric part of the connection, the torsion tensor, becomes a dynamical variable related to the spin density of matter \cite{Hehl}.
Since Dirac fields couple to the affine connection, the intrinsic spin of fermions acts like a source of torsion.
At extremely large densities, existing in black holes and in the very early Universe, the minimal spinor-torsion coupling manifests itself as gravitational repulsion, which avoids the formation of singularities from fermionic matter \cite{avert}.
Accordingly, the big bang is replaced by a nonsingular bounce, before which the Universe was contracting \cite{boun}.
In addition to eliminating the initial singularity, this scenario solves the flatness and horizon problems \cite{alt}.
The ECSK theory therefore provides the simplest and most natural mechanism that solves the above three major problems of the standard big-bang cosmology, without introducing additional matter fields or specific conditions on their form.
The ECSK gravity also passes all tests of general relativity, because even at nuclear densities the contribution from torsion to the Einstein equations is negligibly small and both theories give indistinguishable predictions.

The theory of cosmic inflation also solves the flatness and horizon problems, but it does not address the big-bang singularity \cite{infl}.
It predicts, however, the observed nearly scale-invariant spectrum of cosmological perturbations that are responsible for large-scale structure of the Universe.
These perturbations are formed during the inflationary phase from the initial quantum vacuum fluctuations \cite{pert}.
Such a spectrum can also emerge in a bouncing cosmology, if the Universe in the collapsing phase was dominated by nonrelativistic matter \cite{non}.
Scale-invariant cosmological perturbations after the bounce can also arise from thermal fluctuations in the cold, contracting Universe long before the bounce \cite{the}.

The model of the Universe contracting from infinity in the past does not, however, explain what caused the contraction.
An interesting scenario that solves this problem assumes that such a contraction corresponds to gravitational collapse of matter inside a newly formed black hole existing in another universe \cite{alt,Pat}.
Extremely strong, anisotropic gravitational forces in a collapsing black hole cause an intense pair production, which generates a large amount of mass inside the black hole and isotropizes spacetime in the black hole \cite{Zel}.
At a torsion-induced nonsingular bounce, the interior of a black hole is isotropic and the pair production ceases.
After the bounce, a new (closed) universe in a black hole expands, although such an expansion is not visible from the outside of the black hole due to infinite redshift at its event horizon.
Thermal fluctuations of matter near the bounce are thus the initial condition for the cosmological perturbations in this universe.
Such fluctuations in the ECSK bouncing cosmology \cite{alt,bb} are the subject of this paper.

Before the bounce, all perturbation modes are within the cosmological horizon: their physical wavelengths are smaller than the Hubble radius \cite{the}.
Shortly after the bounce, these modes exit the horizon: their wavelengths exceed the Hubble radius.
Upon the Hubble radius crossing, the matter fluctuations freeze out and the dynamics of the perturbations is governed by the metric.
The linearized fluctuations about a Friedman-Lema\^{i}tre-Robertson-Walker background metric in longitudinal gauge are given by $ds^2=a^2(\eta)[(1+2\Phi)d\eta^2-(1-2\Phi)d{\bf x}^2]$, where $a$ is the background scale factor of the Universe, $\eta$ is the conformal time defined through $d\eta=dt/a(t)$, and $\Phi(\eta,{\bf x})$ is the gauge invariant Bardeen potential (generalized Newtonian gravitational potential) describing the metric fluctuations ($c=\hbar=k_\textrm{B}=1$) \cite{pert}.
The gravitational potential in Fourier space, $\Phi_k$, satisfies an equation
\begin{equation}
\Phi''_k+2\sigma\mathcal{H}\Phi'_k+\Bigl(c^2_s k^2-2(\epsilon-\sigma)\mathcal{H}^2\Bigr)\Phi_k=0,
\label{perturb}
\end{equation}
where $k$ is the comoving wavenumber of the perturbation, $\mathcal{H}=a'/a$ is the comoving Hubble parameter, $c_s$ is the speed of sound, $\epsilon=-\dot{H}/H^2$, $\sigma=-\ddot{H}/(2H\dot{H})$, $H=\dot{a}/a$ is the Hubble parameter, prime denotes differentiation with respect to $\eta$, and dot denotes differentiation with respect to the cosmic time $t$ \cite{the}.
At scales much larger than the Hubble radius $H^{-1}$, which are relevant for the study of cosmological perturbations, the last term on the left side of (\ref{perturb}) is negligibly small.

The time-time component of the perturbed Einstein equations,
\begin{equation}
\nabla^2\Phi-3\mathcal{H}(\Phi'+\mathcal{H}\Phi)=4\pi Ga^2\delta\rho,
\end{equation}
where $\delta\rho$ is the fluctuation of the energy density, relates the metric and matter fluctuations.
The power spectrum of the metric perturbations is equal to
\begin{equation}
P_\Phi(k)=\frac{1}{12\pi^2}k^3|\Phi_k|^2=\frac{1}{4M_\textrm{P}^4 H^4}k^3\langle\delta\rho^2_k\rangle=\frac{1}{4M_\textrm{P}^4 H^4}\langle\delta\rho^2\rangle,
\label{power}
\end{equation}
where brackets $\langle\rangle$ denote ensemble averaging, $\delta\rho_k$ is the fluctuation of the energy density in the momentum space, and $M_\textrm{P}$ is the Planck mass \cite{the}.
The Hubble radius crossing is given by a relation $k=aH=\mathcal{H}$.
The fluctuation correlation function in the position space for the energy density at the Hubble radius crossing must be evaluated in a sphere of radius $R(k)$, where $R(k)$ is the physical length corresponding to the comoving momentum scale $k$ \cite{the}.

In thermal equilibrium, the energy-density correlation function is given by
\begin{equation}
\langle\delta\rho^2\rangle|_{R(k)}=C_V(R)\frac{T^2}{R^6},
\end{equation}
where $C_V(R)=\partial\langle E\rangle/\partial T$ is the heat capacity in a sphere of radius $R$, $T$ is the temperature of the Universe, and $E$ is the internal energy \cite{the}.
The heat capacity is thus
\begin{equation}
C_V(R)=R^3\frac{\partial\rho}{\partial T},
\end{equation}
where $\rho$ is the energy density.
In the Universe near the bounce, the matter is ultrarelativistic: $\rho\sim T^4$.
The power spectrum (\ref{power}) is therefore given by
\begin{equation}
P_\Phi(k)\sim M_\textrm{P}^{-4} H^{-1} T^5.
\label{spect}
\end{equation}

In a closed, homogeneous and isotropic early universe filled with fermionic matter macroscopically averaged as a spin fluid \cite{avert}, the Friedman equations lead to \cite{alt}
\begin{equation}
H=H_0\bigl(\Omega_S\hat{a}^{-6}+\Omega_R\hat{a}^{-4}\bigr)^{1/2},
\label{Hub}
\end{equation}
where $H_0=(a_0 \sqrt{\Omega-1})^{-1}$ is the present Hubble parameter, $a_0$ is the present scale factor, $\hat{a}=a/a_0$ is the normalized scale factor, $\Omega>1$ is the present total density parameter (in the closed Universe), $\Omega_R$ is the present radiation density parameter, and $\Omega_S$ is the negative and extremely small in magnitude, present spinor-torsion density parameter.
The temperature is related to the scale factor through
\begin{equation}
aT\sim \frac{\Omega_R}{\sqrt{\Omega_S(1-\Omega)}}.
\label{temp}
\end{equation}
Since $\Omega_S<0$, the spinor-torsion coupling generates gravitational repulsion which is significant at very small $\hat{a}$, preventing the cosmological singularity $\hat{a}=0$.
The expansion of the Universe starts when $H=0$, at which the normalized scale factor has a minimum but finite value $\hat{a}=\hat{a}_m$:
\begin{equation}
\hat{a}_m=\biggl(-\frac{\Omega_S}{\Omega_R}\biggr)^{1/2}.
\label{mini}
\end{equation}

Replacing the cosmic time $t$ by the conformal time $\eta$ changes (\ref{Hub}) into an equation for $\hat{a}(\eta)$:
\begin{equation}
\frac{d\hat{a}}{(\Omega_S\hat{a}^{-2}+\Omega_R)^{1/2}}=\frac{d\eta}{(\Omega-1)^{1/2}}.
\label{confo}
\end{equation}
If we choose $\eta=0$ at the minimum scale factor, $a(0)=a_m$, then integrating (\ref{confo}) gives
\begin{equation}
a(\eta)=a(0)\biggl(1+\frac{\eta^2}{\eta^2_S}\biggr)^{1/2},
\label{dyn}
\end{equation}
where
\begin{equation}
\eta_S=\frac{\sqrt{\Omega_S(1-\Omega)}}{\Omega_R}
\end{equation}
is the characteristic conformal time at which radiation begins to dominate over the spinor-torsion coupling.
The corresponding Hubble parameter $H=-d(a^{-1})/d\eta$ is equal to
\begin{equation}
H(\eta)=\frac{\eta}{a_0\eta^2_S}\biggl(1+\frac{\eta^2}{\eta^2_S}\biggr)^{-3/2}.
\label{param}
\end{equation}
Equation (\ref{dyn}) represents a two-component fluid composed of radiation and stiff matter.
The perturbations arising from quantum vacuum fluctuations are nearly scale-invariant if the Universe filled with such a fluid has a slowly contracting phase with a different $a(\eta)$ \cite{st}.

The condition for the Hubble radius crossing becomes $k=d\,\textrm{ln}\,a/d\eta$.
For the dynamics (\ref{dyn}), this condition gives
\begin{equation}
\eta^2-k^{-1}\eta+\eta^2_S=0.
\label{cro}
\end{equation}
Because of the intense pair production in the contracting phase of the Universe (in a black hole), we take the thermal fluctuations at the bounce as the initial condition for the subsequent cosmological perturbations.
The solution of (\ref{cro}) at the Hubble radius exit is thus
\begin{equation}
\eta(k)=\frac{1}{2k}\Bigl(1-\sqrt{1-4\eta^2_S k^2}\Bigr),
\end{equation}
showing the conformal time at which the perturbation mode with scale $k$ leaves the horizon.
It also constraints the possible values of $k$: $k\le(2\eta_S)^{-1}$.
At large scales, this solution reduces to $\eta(k)\approx\eta^2_S k$.
Substituting this relation, together with (\ref{temp}), (\ref{dyn}) and (\ref{param}), into (\ref{spect}) gives the power spectrum of the metric perturbations at the horizon exit:
\begin{equation}
P_\Phi(k)\sim(M_\textrm{P} a_0)^{-4}\eta^{-5}_S(1+\eta^2_S k^2)^{-1}k^{-1}\approx(M_\textrm{P} a_0)^{-4}\eta^{-5}_S k^{-1}.
\label{cons}
\end{equation}
The corresponding spectral index $n_S$, defined through $P_\Phi(k)\sim k^{n_S-1}$, is equal to 0, while the scale-invariant spectrum has $n_S=1$.
To compare, taking quantum vacuum fluctuations in the contracting Universe in the remote past as the initial condition for the subsequent cosmological perturbations is even worse: $n_S=-1$ \cite{st}.
The observed spectrum of cosmological perturbations cannot therefore arise from thermal fluctuations at the nonsingular bounce induced by the coupling between the torsion tensor and the fermionic matter approximated as a spin fluid in the ECSK gravity \cite{avert,alt}.

The spin-fluid description of macroscopic fermionic matter, however, is based on the particle approximation of Dirac fields, which is not consistent with the conservation law for the spin tensor \cite{nonsi}.
In \cite{bb}, we used the Dirac spin tensor, which follows from the Dirac Lagrangian for fermions without any approximations.
We showed that the minimal coupling between the torsion tensor and fermionic matter with the Dirac form of the spin tensor also leads to gravitational repulsion at extremely high densities on the same order as those for spin fluids.
Such a repulsion avoids the initial singularity and solves the flatness and horizon problems \cite{bb}, as in \cite{infl}.
While the bounce generated by the spin fluid coupled to torsion is characterized by $\dot{a}=0$, which very rapidly grows to an enormous quantity (the Hubble radius rapidly decreases from infinity to a very small quantity and perturbation modes exit the horizon), the bounce generated by the Dirac spinors coupled to torsion is different: $\dot{a}$ jumps there from $\dot{a}=-v$ to $\dot{a}=v$, where $v$ is a very large quantity \cite{bb}.
Accordingly, while perturbation modes in the former scenario leave the horizon at different conformal times that depend on $k$, in the latter they leave the horizon at the same conformal time: the conformal time of the bounce.
The resulting power spectrum of the metric perturbations at the horizon exit does not depend on $k$.
The ECSK gravity combined with the Dirac theory of fermions therefore not only provides the simplest and most natural mechanism that solves all major problems of the standard cosmology, but also predicts the scale-invariant spectrum of cosmological perturbations.
The implications of the Einstein-Cartan-Sciama-Kibble-Dirac bouncing cosmology on the magnitude of the density inhomogeneities, $\delta\rho/\rho$, will be investigated subsequently.


\begin{thebibliography}{}
\bibitem{KS} T. W. B. Kibble, J. Math. Phys. (N.Y.) {\bf 2}, 212 (1961); D. W. Sciama, in {\em Recent Developments in General Relativity} (Pergamon, Oxford, 1962), p. 415; D. W. Sciama, Rev. Mod. Phys. {\bf 36}, 463 (1964); D. W. Sciama, Rev. Mod. Phys. {\bf 36}, 1103(E) (1964).
\bibitem{Hehl} F. W. Hehl and B. K. Datta, J. Math. Phys. (N.Y.) {\bf 12}, 1334 (1971); F. W. Hehl, Phys. Lett. A {\bf 36}, 225 (1971); F. W. Hehl, Gen. Relativ. Gravit. {\bf 4}, 333 (1973); F. W. Hehl, Gen. Relativ. Gravit. {\bf 5}, 491 (1974); F. W. Hehl, P. von der Heyde, G. D. Kerlick, and J. M. Nester, Rev. Mod. Phys. {\bf 48}, 393 (1976); E. A. Lord, {\em Tensors, Relativity and Cosmology} (McGraw-Hill, New Delhi, 1976); V. de Sabbata and M. Gasperini, {\em Introduction to Gravitation} (World Scientific, Singapore, 1985); V. de Sabbata and C. Sivaram, {\em Spin and Torsion in Gravitation} (World Scientific, Singapore, 1994); I. L. Shapiro, Phys. Rep. {\bf 357}, 113 (2002); R. T. Hammond, Rep. Prog. Phys. {\bf 65}, 599 (2002); N. J. Pop{\l}awski, arXiv:0911.0334.
\bibitem{avert} W. Kopczy\'{n}ski, Phys. Lett. A {\bf 39}, 219 (1972); W. Kopczy\'{n}ski, Phys. Lett. A {\bf 43}, 63 (1973); A. Trautman, Nature (Phys. Sci.) {\bf 242}, 7 (1973); J. Tafel, Phys. Lett. A {\bf 45}, 341 (1973); F. W. Hehl, P. von der Heyde, and G. D. Kerlick, Phys. Rev. D {\bf 10}, 1066 (1974); B. Kuchowicz, Gen. Relativ. Gravit. {\bf 9}, 511 (1978); M. Gasperini, Phys. Rev. Lett. {\bf 56}, 2873 (1986); M. Gasperini, Gen. Relativ. Gravit. {\bf 30}, 1703 (1998).
\bibitem{boun} M. Novello and S. E. P. Bergliaffa, Phys. Rept. {\bf 463}, 127 (2008); M. Lilley, L. Lorenz, and S. Clesse, J. Cosm. Astropart. Phys. {\bf 06}, 004 (2011).
\bibitem{alt} N. J. Pop{\l}awski, Phys. Lett. B {\bf 694}, 181 (2010); N. J. Pop{\l}awski, Phys. Lett. B {\bf 701}, 672(E) (2011); N. J. Pop{\l}awski, Gen. Relativ. Gravit., in press (2012), arXiv:1105.6127.
\bibitem{infl} D. Kazanas, Astrophys. J. {\bf 241}, L59 (1980); A. H. Guth, Phys. Rev. D {\bf 23}, 347 (1981); A. Linde, Phys. Lett. B {\bf 108}, 389 (1982).
\bibitem{pert} J. M. Bardeen, Phys. Rev. D {\bf 22}, 1882 (1980); V. F. Mukhanov, H. A. Feldman and R. H. Brandenberger, Phys. Rept. {\bf 215}, 203 (1992).
\bibitem{non} D. Wands, Phys. Rev. D {\bf 60}, 023507 (1999).
\bibitem{the} Y.-F. Cai, W. Xue, R. Brandenberger, and X. Zhang, J. Cosm. Astropart. Phys. {\bf 06}, 037 (2009).
\bibitem{Pat} R. K. Pathria, Nature {\bf 240}, 298 (1972); V. P. Frolov, M. A. Markov, and V. F. Mukhanov, Phys. Lett. B {\bf 216}, 272 (1989); V. P. Frolov, M. A. Markov, and V. F. Mukhanov, Phys. Rev. D {\bf 41}, 383 (1990); L. Smolin, Class. Quantum Grav. {\bf 9}, 173 (1992); W. M. Stuckey, Am. J. Phys. {\bf 62}, 788 (1994); D. A. Easson and R. H. Brandenberger, J. High Energy Phys. {\bf 06}, 024 (2001); J. Smoller and B. Temple, Proc. Natl. Acad. Sci. USA {\bf 100}, 11216 (2003); N. J. Pop{\l}awski, Phys. Lett. B {\bf 687}, 110 (2010).
\bibitem{Zel} L. Parker, Phys. Rev. {\bf 183}, 1057 (1969); Ya. B. Zel'dovich, J. Exp. Theor. Phys. Lett. {\bf 12}, 307 (1970); Ya. B. Zel'dovich and A. A. Starobinskii, J. Exp. Theor. Phys. Lett. {\bf 26}, 252 (1978).
\bibitem{bb} N. J. Pop{\l}awski, arXiv:1111.4595.
\bibitem{st} P. Peter and N. Pinto-Neto, Phys. Rev. D {\bf 66}, 063509 (2002).
\bibitem{nonsi} N. J. Pop{\l}awski, Phys. Lett. B {\bf 690}, 73 (2010).
\end{thebibliography}
\end{document}